\documentclass[aps,pra,twocolumn,showpacs,superscriptaddress,groupedaddress]{revtex4} 

\usepackage{graphicx} 
\usepackage{dcolumn}
\usepackage{bm}
\usepackage{amssymb}
\usepackage{amsmath}
\usepackage{hyperref}

\begin{document}

\title{Multiphoton inner-shell ionization of the carbon atom}
\author{H.F. Rey}
\email{hreypereira01@qub.ac.uk}
\author{H.W. van der Hart}
\affiliation{Centre for Theoretical Atomic, Molecular and Optical Physics, School
of Mathematics and Physics, Queen's University Belfast, Belfast BT7 1NN, United Kingdom}

\begin{abstract}
We apply time-dependent R-matrix theory to study inner-shell ionization of C atoms
in ultra-short high-frequency light fields
with a photon energy between 170 and 245 eV. At an intensity of 10$^{17}$ W/cm$^2$,
ionization is
dominated by single-photon emission of a $2\ell$ electron, with two-photon emission
of a 1s electron accounting for 
about 2-3\% of all emission processes, and two-photon emission of $2\ell$ contributing
about 0.5-1\%. Three-photon emission of a 1s electron is estimated to contribute about
0.01-0.03\%. Around a photon energy of 225 eV, two-photon emission of a 1s electron,
leaving C$^+$ in either 1s2s2p$^3$ or 1s2p$^4$ is resonantly enhanced by intermediate
1s2s$^2$2p$^3$ states. The results demonstrate the capability of time-dependent R-matrix
theory to describe inner-shell ionization processes including rearrangement of the outer
electrons.
\end{abstract}


\maketitle

\section{Introduction}

Over the last decade, great strides have been made in the development of free-electron
lasers
operating in the VUV - X-ray regime. Several free-electron lasers operating in the
XUV-X-ray regime have
become available to the community in recent years: for example, FLASH \cite{Flash-JPB},
LCLS \cite{LCLS-JPB}, and
SACLA \cite{Japan-JPB}. These
facilities have demonstrated their potential for opening new areas of atomic, molecular
and optical physics, for
example through the study
of Auger resonances which cannot be excited by a single photon \cite{Flash-exp},
multiphoton sequential ionization of Xe up to Xe$^{36+}$ at a photon energy of 1.5 keV
 \cite{LCLS-exp} and multiphoton multiple ionization of N$_2$ \cite{Japan-exp}. 

Photoionization in high-frequency laser fields tends to be dominated by the innermost
electron that can be ejected.
However, the outer electrons will also experience the light field, and can therefore
still absorb a photon. Hence, a full description
of the atomic or molecular response should consider all electrons that could
possibly be affected by the laser field.
In addition, outer electrons do not necessarily remain in their original orbital when an
inner electron is removed from the system.
The potential seen by the outer electrons may change suddenly, leading to shake-up
excitation of the outer electrons. 

A full theoretical or computational study of the interaction between high-frequency
laser light and atoms
will therefore require a method which can describe the simultaneous
response of many electrons to the laser field. Several
such methods have been developed in recent years, such as the time-dependent
configuration-interaction 
singles (TDCIS) method, which has recently been applied to study above-threshold ionization for light elements in
the hard X-ray regime \cite{Til15}, a Green's function technique, algebraic diagrammatic construction (ADC), which
has been applied to study fast dynamics in glycine using laser pulses at a photon energy of 275 eV \cite{Coo14}, and
time-dependent R-matrix theory, which has been used to study the competition between emission of a 2s and a
2p electron in C in the UV regime \cite{Rey14}.

In the present study, we continue our study of the response of  C atoms to laser light by investigating the
photon energy range between 170 and 245 eV. The response of the carbon atom is of particular interest as
it is the prime constituent of biological molecules. The removal of inner electrons from a
carbon atom can provide new insight
into molecular systems. For example, in \cite{Ple12}, it was demonstrated that the removal of a 1s electron from the
carbon atom in methane could be exploited experimentally to extract information about the molecular geometry. In
\cite{Coo14}, it was also proposed that dynamics in ionized glycine could be studied in a pump-probe scheme where the probe
pulse excites a localised 1s electron of C to the orbital in which a hole is created by the pump pulse. A photon energy
of 275-280 eV was suggested for this purpose. Advances in laser technology have very
recently been exploited
to generate individual subfemtosecond pulses in this photon energy range \cite{Sil15}.
Thus it is of interest to investigate ultra-fast dynamics involving
inner-shell electrons.

As a first step towards the treatment of short laser pulses at a photon energy of 284
eV \cite{Sil15}, we
compare in the present study
emission of the inner 1s electron with emission of the outer 2s or 2p electrons
for the photon energy range
between 170 and 245 eV. Numerical studies of ultra-fast dynamics at 284 eV involving
the C atom will require great care with the pulse shape to
ensure
that inner-shell ionization processes are not dominated by direct single-photon emission
arising from the outer edges
of the pulse bandwidth. In the present photon energy range, emission of a 1s electron
requires absorption of (at least)
two photons, whereas the emission of a 2s or 2p electron requires absorption of a single photon only. This comparison
is similar to a previous comparison of two-photon emission of the 1s electron versus
one- and two-photon emission of
the outer 2s electron using the R-matrix Floquet approach for Li$^-$ \cite{Har05} or
the 1s2s $^1$S state in He \cite{Mad06}. 

To study the response of the carbon atom, we use the recently developed time-dependent R-matrix approach
RMT \cite{Nik08,Lys11,Moo11}. It combines the capability of R-matrix theory to describe a wide range of processes in
general atomic systems \cite{Har07, Gua07,Bur11} with the computational capability of the
HELIUM approach \cite{Smy98}. The combination
of these two techniques has enabled the determination of time delays in Ne \cite{Moo11}, and
 high-harmonic
generation at mid-IR wavelengths \cite{Has14}. For these studies, relatively little atomic
structure was taken into account. By adopting an R-matrix with pseudo-states philosophy \cite{Bar96}
or an intermediate-energy R-matrix approach \cite{Sco09}, we have recently demonstrated that the RMT codes can
also be used for the study of double photoionization processes \cite{Har14,Wra15}. To describe the double continuum
accurately, extensive atomic structure needs to be taken into account. The success of these latter studies suggests
that the RMT approach is capable of treating atoms in strong fields with
a detailed description of atomic structure.

In the present study, we wish to explore the application of RMT theory to a case where electrons can be ejected from
different shells with significantly different binding energy. This application poses new demands on the computational
accuracy. The continuum needs to be accurate up to very high energies to describe relevant above-threshold ionization
processes involving outer electrons. This has the computational consequence of increasing
the round-off error in
matching the wavefunction at the boundary between the inner and outer R-matrix regions, where the description of
the ejected electron changes from basis-set techniques to finite-difference methods. Hence the application
of the RMT approach to inner-shell processes presents new demands on the computer codes.

In section \ref{sec:comp}, we give a short overview of the RMT approach. We also provide a brief description
of the basis set used to describe the C atom, and the laser pulse. The results are presented in section \ref{sec:results}
with an emphasis on the competition between the multiphoton emission of an inner 1s electron
and single- and
multiphoton emission of an outer $2\ell$ electron. 

\section{Computational methods}
\label{sec:comp}

Time-dependent R-matrix theory is the extension of the standard R-matrix approach \cite{Bur11}
to the solution of the time-dependent Schr\"odinger equation \cite{Har07, Gua07, Nik08,
Lys09, Moo11, Lys11}. Although the initial applications of time-dependent R-matrix
theory described electrons restricted to a finite region surrounding the nucleus
\cite{Har07, Gua07}, subsequent implementations adopted the standard R-matrix concept
of division of space into two distinct regions: an inner region in which all electrons
are close to the nucleus, and an outer region in which one electron has moved well
away from the others. In the original formulation of time-dependent R-matrix theory,
an R-matrix propagation scheme was employed to propagate the wavefunction \cite{Lys09}.
This approach relies on the solution of systems of equations throughout the calculation,
and as a consequence calculation time increases rapidly with an increase in atomic
structure.

The most recent implementation of time-dependent R-matrix theory is R-matrix theory
with time dependence (RMT) \cite{Moo11, Lys11}. In this approach, the wavefunction
in the inner region is described in terms of a standard R-matrix basis with a
B-spline representation of the continuum orbitals. The wavefunction in the outer
region is described in terms of a direct product of a residual-ion state coupled
with a finite-difference representation of the wavefunction for the outer electron.
Near the boundary between inner and outer region, the wavefunction must be shared by
the inner and outer regions. This is achieved through evaluation of the inner-region
wavefunction on an outer-region grid extension into the inner region for use by the
outer region. The outer-region wavefunction information needed by the inner region
consists of spatial derivatives of the outer-region wavefunction at the inner-region
boundary.

The main advantage of the RMT approach over the previous implementation is its
improved accuracy and numerical efficiency. Whereas the previous
implementation used a low-order Crank-Nicolson propagator, the RMT approach uses a
high-order Arnoldi propagator \cite{Smy98}. This replaces a solution of a system of linear equations
by repeated matrix-vector multiplications, which may reduce numerical error in the
time and spatial propagation of the wavefunction. Since the Arnoldi propagator is
dominated by matrix-vector multiplications, the RMT codes can be parallelised more
efficiently, so that calculations exploiting in excess of 2000 cores are feasible. 

In the present study, we aim to investigate inner-shell ionization processes involving
the carbon atom. We are thus interested in residual-ion states with a hole in the 1s
shell. Within the R-matrix codes, residual-ion states are retained in order of energy.
As a consequence, all possible residual-ion states with a filled 1s$^2$ shell are
included prior to inclusion of the residual-ion states with a hole in the 1s shell.
In order to limit the scale of the calculations, we therefore adopt a minimal basis
for the description of carbon. The atom is described using only the 1s, 2s and 2p
Hartree-Fock orbitals of singly ionized carbon \cite{Cle74}. We then use these
orbitals to build all possible singly-charged residual-ion states: i.e. all states
belonging to the 1s$^2$2s$^2$2p, 1s$^2$2s2p$^2$, 1s$^2$2p$^3$, 1s2s$^2$2p$^2$,
1s2s2p$^3$ and 1s2p$^4$ configurations of C$^+$. The neutral-atom basis then
contains all combinations of these residual-ion states with a set of B-spline
based continuum orbitals up to a maximum total angular momentum $L_{\rm max} = 5$. We use a
total of 125 B-splines of order 9 to build these continuum orbitals. This basis also
includes so-called correlation functions made of all combinations for 6 electrons
across the 1s, 2s and 2p orbitals with at least one electron in 1s. The inner-region
boundary is set to 27 $a_0$.

The application of RMT theory using this basis set to describe the atom poses a new
challenge: a challenge that applies
to all studies of inner-shell ionization processes. To describe all ionization
processes properly, a good description of the continuum is needed for emission of
both inner and outer electrons, including multiphoton emission processes. Multiphoton
emission of outer electrons, in particular, can lead to very high continuum energies.
An extensive expansion of the continuum is therefore required in the inner region,
which allows for the
description of rapidly oscillating continuum functions. As a consequence, the so-called
knot points of the B-spline basis set are more closely spaced than in a calculation for
outer electrons only: 125 B-splines in the present case, compared to 70 for the
outer-electron calculation \cite{Rey14}. The introduction of
the Bloch operator to maintain Hermiticity of the Hamiltonian in the inner region then
generates eigenfunctions with large eigenvalues (up to 200 keV) which are sharply peaked
near the R-matrix boundary. Since these functions are peaked near the boundary, they need
to be retained in the calculations. However, the effect of the large eigenvalues must be
compensated for through the connection between the inner region and the outer region.
Hence, a cancellation of terms involving large energies occurs at every stage of the
calculation. This cancellation can be a prime source of numerical error, and extra care
therefore needs to be taken to ensure numerical stability of the calculations, for example
through a significant reduction of the time-step in the calculation.  

Within the RMT approach, the light field is assumed to be linearly polarized and described
within the length form of the dipole approximation
due to the necessity to restrict the residual-ion basis
\cite{Hut11}. The field is described by an ultra-short light pulse of eight cycles,
including a three-cycle $\sin^2$ ramp-on and ramp-off, with two cycles at peak intensity.
The photon energies in the present study range from 170 eV to 245 eV, so that a single
photon suffices to eject an outer $2\ell$ electron, but absorption of two photons is
required to emit the 1s electron. The bandwidth of the pulse is about 40 eV (full width at half
maximum) at a photon
energy of 245 eV.
After the pulse has ended, we propagate the wavefunction for another 42 cycles
to ensure that all ejected electrons have entered the outer region. The time-step used
in the calculation is 0.012 as. The outer-region finite difference grid has a spacing of
0.025 $a_0$, and extends out to a distance of 816 $a_0$. We use an Arnoldi propagator
of order 10.

\section{Results}
\label{sec:results}

In the present study, we aim to investigate the competition between single-photon emission
of a $2\ell$ electron and two-photon emission of a 1s electron from a carbon atom in the
photon energy range between 170 and 245 eV. Since the C ground state has $^3$P$^e$
symmetry, the initial state can have $M_L=-1, 0$ and 1. In all results presented,
unless otherwise stated, we have averaged over the different initial $M_L$-values.
For non-zero initial $M_L$, the $S$ symmetry is not available, whereas for zero $M_L$
radiative transitions with $\Delta L =0$ are not allowed.

\begin{table*}
\caption{Final-state populations in the outer region for ground-state C atoms irradiated
by an ultra-short laser pulse
with a central photon energy of 190 and 245 eV at different peak laser intensities. The
ground state of C has
even parity. The notation 1.54(-6) indicates 1.54 $\times$ 10$^{-6}$. The populations
are averaged over initial orbital magnetic quantum number $M_L$.}
\begin{tabular}{lc|ccccc}
Channel subset & Photon energy & \multicolumn{5}{c}{Peak intensity} \\
&  eV & 10$^{14}$ W/cm$^2$ & 10$^{15}$ W/cm$^2$ &10$^{16}$ W/cm$^2$ &10$^{17}$ W/cm$^2$ &10$^{18}$ W/cm$^2$  \\ \hline
C$^+$ 2$\ell$ emission, odd parity & 190 & 3.89(-5) & 3.89(-4) & 3.87(-3)& 3.64(-2) & 2.07(-1)\\ 
C$^+$ 2$\ell$ emission, even parity & 190 & 2.45(-10) & 2.45(-8)& 2.45(-6)& 2.45(-4) & 2.68(-2)\\ 
C$^+$ 1s emission, even parity & 190 & 6.59(-10) & 6.59(-8)& 6.55(-6)& 6.18(-4) & 3.66(-2)\\ 
C$^+$ 1s emission, odd parity & 190 & 2.55(-9) & 2.55(-8)& 2.57(-7)& 7.92(-6) & 3.16(-3)\\ \hline
C$^+$ 2$\ell$ emission, odd parity & 245 & 1.29(-5) &1.28(-4) & 1.28(-3)& 1.24(-2) & \\ 
C$^+$ 2$\ell$ emission, even parity & 245 & 5.05(-11) & 5.05(-9)& 5.02(-7)& 4.76(-5) & \\ 
C$^+$ 1s emission, even parity & 245 & 3.19(-10) & 3.18(-8)& 3.16(-6)& 2.96(-4) & \\ 
C$^+$ 1s emission, odd parity & 245 & 3.90(-7) & 3.90(-6)& 3.90(-5)& 3.89(-4) & 
\end{tabular}
\label{tab:seven}
\end{table*}

In table \ref{tab:seven}, we present final-state populations in the outer region for
various subsets of
photoemission channels when carbon is irradiated by a short pulse of 190 eV and 245 eV
photons at various peak laser intensities. The table shows that the yield for
odd-parity channels associated with emission of a 2$\ell$ electrons scales
approximately linearly with intensity between 10$^{14}$ and 10$^{17}$ W/cm$^2$.
This indicates that
these channels correspond to single-photon emission of an outer $2\ell$ electron.
The population in the even-parity channels associated with 2$\ell$ emission increases
quadratically with intensity, and this population can thus be interpreted as two-photon
above-threshold emission of a 2$\ell$ electron. Similarly, the population in the
even-parity channels associated with 1s emission increases quadratically with
intensity, and this population can be interpreted as two-photon emission of a
1s electron. Up to intensities of 10$^{17}$ W/cm$^{-2}$, the assumption that perturbation
theory applies is therefore not unreasonable, with deviations typically less than 7\%.

At an intensity of 10$^{18}$ W/cm$^2$, the assumption of perturbation theory certainly no longer
holds. The single-photon emission yield for a $2\ell$ electron increases by about a factor
6 between 10$^{17}$ W/cm$^2$. The two-photon emission yield for a 1s electron increases
by about a factor 60 between 10$^{17}$ W/cm$^2$. In both cases, a reduction of about 40\%
from the perturbative result is observed. For two-photon emission of a $2\ell$ electron,
an additional increase in the yield of about 10\% is seen. Total ionization has become
substantial at 10$^{18}$ W/cm$^2$, with about 27\% of all atoms ionized within a short time.
Hence, perturbation should no longer be expected to apply at this intensity. Perturbation
theory could fail in different respects: an increase in importance of higher-order
processes, and saturation of ionization, in combination with the very short duration of
the XUV pulse (smaller than $2\pi/I_p$ with $I_p$ the ionization potential of the 2p electron).
Also, the strength of the electric field will significantly change the actual potential
seen by the outer electrons.

Table \ref{tab:seven} shows, furthermore, that the population in odd-parity channels
associated with 1s emission scales linearly with intensity at the lowest
intensities for 190 eV photons and at all intensities for 245 eV photons. The reason
for this is the large bandwidth of the ultra-short light pulse. The threshold for
single-photon emission of the 1s electron in the present calculations is about 300
eV. (This threshold is not necessarily determined very accurately, since the lowest
1s emission threshold, 1s2s$^2$2p$^2$ $^4$P$^e$, is described using Hartree-Fock
orbitals for the 1s$^2$2s$^2$2p ground state of C$^+$.) The ultra-short nature of
the light pulse has sufficient bandwidth  to allow single-photon ionization of a
1s electron to occur with a relative strength, compared to single-photon emission
of a $2\ell$ electron, of about 10$^{-4}$ at 190 eV and about 3$\times$10$^{-2}$
at 245 eV. This increase with increasing photon energy is expected, as a larger
photon energy will lead to a larger part of the bandwidth having sufficient energy
to eject a 1s electron. At 245 eV, this bandwidth leads to dominance of single-photon
emission of a 1s electron over multiphoton emission even at an intensity of 10$^{18}$
W/cm$^2$. However, at 190 eV, the population of the 1s emission channels with odd
parity increases by a factor of 400 when the intensity is increased from 10$^{17}$
W/cm$^2$ to 10$^{18}$ W/cm$^2$. This indicates that for this photon energy,
three-photon processes start to become important near an intensity of 10$^{17}$
W/cm$^2$. We can therefore obtain an estimate for three-photon emission of a 1s electron
through comparison of the final-state populations obtained at intensities of 10$^{14}$
W/cm$^2$ and 10$^{17}$ W/cm$^2$. This comparison gives a final-state population in 1s
emission channels associated with three-photon absorption at an intensity of 10$^{17}$
W/cm$^2$ of around 5.4 $\times$ 10$^{-6}$. This procedure to estimate above threshold
emission for the 1s electron can be carried out for photon energies up to about 220 eV.

The final-state populations in table \ref{tab:seven} have been averaged over $M_L$.
Little difference is seen between the final-state populations for $M_L=0$ and
$M_L=\pm1$, except for two-photon emission of the 2$\ell$ electrons, for which
the $M_L=\pm1$ yield is about 65\% larger than the $M_L=0$ population
at 190 eV, a factor 2 larger at 218 eV and 10\% smaller at 245 eV. The most likely
reason for the generally larger yield for $M_L=\pm1$ is the presence of a 2p electron
with $m_\ell=0$ for the initial $M_L=1$ level of the $^3$P$^e$ ground state of C,
whereas no such electron is present for the initial $M_L=0$ level. We note that
this same principle
was the reason for the relative increase in the emission of 2s electrons compared
to 2p electrons in multiphoton ionization of C at 390 nm for $M_L=0$ \cite{Rey14}.

We can compare the ionization yields in the odd-parity channels with estimates of the
photoionization cross sections in the various subshells \cite{Yeh85}. The calculation
we compare with also uses a very simple description for C. First we consider the results
for a central photon energy of 190 eV. The ionization yield for $2\ell$
electrons gives a photoionization cross section for the $n=2$ subshell of 0.128 Mb, very
similar to the result presented in \cite{Yeh85}. The determination of a 1s photoionization
cross section is more difficult. The Fourier transform of the electric field shows that
the photon energy spectrum above the 1s ionization threshold accounts for about 2.3 $\times$
$10^{-5}$ of the full intensity. It contains three main components:
20\% of the intensity associated with photon energy spectrum above the 1s ionization threshold
is found within 20 eV of the threshold, 70\% of the intensity is associated
with photon energies around 320 eV, and 10\% with photon energies around 365 eV. If
we assume that the full 1s emission yield arises from photon energies
around 320 eV, a cross section of 0.64 Mb is obtained, which compares
reasonably  well with a reported
photoionization cross section of 0.86 Mb at a photon energy of 300 eV \cite{Yeh85}, given the
various uncertainties in transforming the yield into a cross section.

Using the same procedure, we obtain a photoionization cross
section of 0.070 Mb for the $n=2$ subshell for
a central photon energy of 245 eV, in good agreement with the result reported
in \cite{Yeh85}. The estimate of a 1s photoionization cross section is
again more complicated. The Fourier transform of the electric field shows that
the photon energy spectrum above the 1s ionization threshold accounts for 0.88\% of the
full intensity. It contains two main components:
33\% of the intensity with photon energy above the 1s ionization threshold is found
within 8 eV of the threshold, with the remaining 67\% of the
intensity associated with photon energies peaking around 310 eV. If
we assume that the full 1s emission yield arises from photon energies
around 310 eV, a cross section of 0.31 Mb is obtained, which is nearly a factor 3 smaller
than the reported photoionization cross section of 0.86 Mb at a photon energy of 300 eV \cite{Yeh85}.
The reason for this increase in discrepancy is unknown. The components near threshold are strongest
at threshold, and may therefore not lead to ionization within the calculations, due to the
time needed by the photoelectron
to leave the inner region. However, this would only increase our estimate of the
cross section to 0.46 Mb, which is still notably below the previous result.
Other possible reasons for the discrepancy include the short duration of the pulse, and
possible interference arising from excitation of the 1s electron to $n$p states by the
central component of the pulse.

\begin{figure}
\includegraphics[width=9cm]{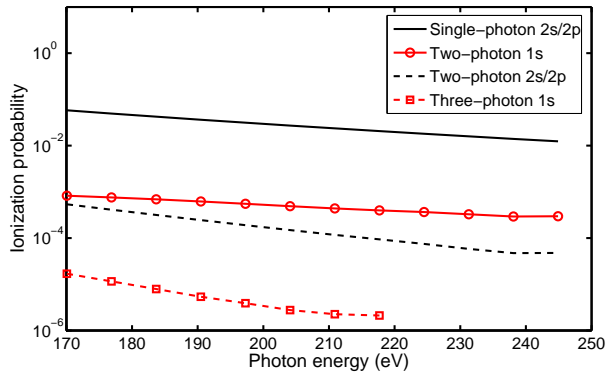}
\caption{(Colour online) Yields for one- and two-photon emission of a $2\ell$ electron and
two- and three-photon emission of a 1s electron from neutral C irradiated by
a short laser pulse with a peak intensity of 10$^{17}$ W/cm$^2$ as a function
of photon energy. The three-photon emission yield is estimated through subtraction
of the single-photon yield induced by the edges of the pulse bandwidth. Data associated
with this report can be accessed via \cite{Pure}.}
\label{fig:carbonemit}
\end{figure}

Table \ref{tab:seven} shows that, through consideration of the outer-region channels,
we can obtain emission yields for single-photon emission of a 2s or 2p electron,
two-photon emission of a 1s electron, two-photon emission of a 2s or 2p electron,
and an approximation to three-photon emission yield of a 1s electron at lower photon
energies at an intensity of 10$^{17}$ W/cm$^2$. 
Figure \ref{fig:carbonemit} shows these emission yields as a function of photon energy.
The figure demonstrates that multiphoton emission is not negligible at this intensity,
as it contributes between 2 and 3\% to the total emission. The figure
demonstrates that two-photon emission of the inner 1s electron is more important than
two-photon emission of an outer 2s or 2p electron, and the importance of two-photon
emission of a 1s electron increases with photon energy, compared to both single-photon
and two-photon emission of a $2\ell$ electron. Three-photon emission of the 1s electron
contributes about 0.02\% to the total photoemission yield for a photon energy of 170 eV
and an intensity 10$^{17}$ W/cm$^2$. This contribution drops to about 0.01\% at a photon
energy of 210 eV.

At the highest intensity considered in the present calculation, 10$^{18}$ W/cm$^2$,
the increase in ionization yield no longer follows a perturbative scaling law. However,
the relative importance of higher-order processes continues to increase. At this
intensity, two-photon emission of a 1s electron accounts for 16\% of all photoemission
processes, with two-photon emission of a $2\ell$ electron accounting for another 2\%.
Therefore at intensities approaching 10$^{18}$ W/cm$^2$, a full analysis of the physics
must account for the potential effects of absorption of more than just a single photon. 

Previously, we investigated the competition between multiphoton absorption by a 1s
and 2s electron from the initial 1s2s $^1$S
state in He \cite{Mad06}. In that study, it was found that at the two-photon level,
the 1s electron was about five times as likely as the 2s
electron to absorb two photons. In the present study, at 175 eV, the 1s electron is
about 2.5 times as likely to absorb two photons than for a
$2\ell$ electron to absorb two photons, whereas it is a factor 6 at a photon energy
of 245 eV. The ratio obtained here is therefore comparable to the one found for He,
even though in the present case, there are twice as many $2\ell$ electrons as $1s$
electrons. On the other hand, as shown later, the emission of $2\ell$ electrons is
dominated by emission of a 2s electron, with 2p electrons contributing notably less.

\begin{table}
\caption{Populations of final states of C$^+$ after irradiation of a ground-state C
atom by an eight-cycle laser pulse with an intensity of 10$^{17}$ W/cm$^2$  at photon
energies of 170 eV, 197 eV and 224 eV. The notation 1.16(-2) denotes 1.16 $\times$ 10$^{-2}$.
}
\begin{tabular}{l|rrr}
C$^+$ & \multicolumn{3}{c}{Final-state population} \\
& \multicolumn{3}{c}{at a photon energy of}\\
state & 170 eV & 197 eV & 224 eV\\ \hline
1s$^2$2s$^2$2p $^2$P$^o$ & 1.16(-2) & 5.35(-3) & 2.64(-3)\\ \hline
1s$^2$2s2p$^2$ $^4$P$^e$ & 3.07(-2) & 1.70(-2) & 9.98(-3)\\
1s$^2$2s2p$^2$ $^2$D$^e$ & 4.92(-4) & 2.12(-4) & 1.11(-4)\\
1s$^2$2s2p$^2$ $^2$S$^e$ & 7.47(-5) & 3.43(-5) & 2.03(-5)\\
1s$^2$2s2p$^2$ $^2$P$^e$ & 1.24(-2) & 7.24(-3) & 4.43(-3)\\ \hline
1s$^2$2p$^3$ $^4$S$^o$ & 9.92(-4) & 5.07(-4) & 2.88(-4)\\
1s$^2$2p$^3$ $^2$D$^o$ & 1.55(-3) & 7.81(-4) & 4.29(-4)\\
1s$^2$2p$^3$ $^2$P$^o$ & 8.29(-4) & 4.40(-4) & 2.51(-4)\\ \hline
1s2s$^2$2p$^2$ $^4$P$^e$ & 4.90(-4) & 3.56(-4) & 2.29(-4) \\
1s2s$^2$2p$^2$ $^2$P$^e$ & 2.91(-4) & 1.74(-4) & 1.08(-4) \\
1s2s$^2$2p$^2$ $^2$D$^e$ & 9.27(-6) & 8.80(-7) & 2.52(-6) \\
1s2s$^2$2p$^2$ $^2$S$^e$ & 3.18(-6) & 1.61(-7) & 3.58(-7) \\ \hline
1s2s2p$^3$ $^4$S$^o$ & 2.71(-6) & 1.21(-6) & 5.01(-6) \\
1s2s2p$^3$ $^4$D$^o$ & 9.94(-6) & 8.06(-6) & 8.43(-6) \\
1s2s2p$^3$ $^4$P$^o$ & 4.47(-6) & 3.60(-6) & 4.60(-6) \\
1s2s2p$^3$ $^2$D$^o$ & 1.12(-5) & 4.66(-6) & 2.85(-6) \\
1s2s2p$^3$ $^2$P$^o$ & 5.46(-6) & 1.85(-6) & 1.30(-6) \\
1s2s2p$^3$ $^2$D$^o$ & 3.51(-6) & 1.03(-6) & 2.63(-6) \\
1s2s2p$^3$ $^4$S$^o$ & 4.29(-6) & 3.03(-6) & 1.80(-6) \\
1s2s2p$^3$ $^2$S$^o$ & 1.62(-6) & 1.23(-6) & 2.48(-6) \\
1s2s2p$^3$ $^2$P$^o$ & 3.11(-6) & 5.17(-7) & 1.49(-6) \\ \hline
1s2p$^4$ $^4$P$^e$ & 4.59(-7) & 2.85(-7) & 3.37(-7) \\
1s2p$^4$ $^2$D$^e$ & 7.30(-7) & 1.79(-7) & 3.48(-7) \\
1s2p$^4$ $^2$S$^e$ & 3.48(-7) & 7.21(-8) & 2.09(-7) \\
1s2p$^4$ $^2$P$^e$ & 2.11(-7) & 3.17(-8) & 3.49(-8) 
\end{tabular}
\label{tab:ionstatepop}
\end{table}

The RMT calculations also provide the populations in the various residual C$^+$ states.
Table \ref{tab:ionstatepop} provides the final populations for all C$^+$ states included
in the calculations for photon energies of 170 eV, 197 eV and 224 eV at an intensity of
10$^{17}$ W/cm$^2$. For the photon energies given in the table, effects due to the
bandwidth of the pulse are expected to play only a minor role, and should not affect
the final populations significantly.

The table demonstrates that the main residual ion states are the
1s$^2$2s2p$^2$ $^4$P$^e$ and $^2$P$^e$ states indicating a dominance of
direct emission of a 2s electron. Emission of a 2p electron accounts for 20\%
of the emission processes at 170 eV and only 14\% at 224 eV. C$^+$ is left in
a 1s$^2$2p$^3$ state in about 5.2-5.7\% of all emission processes. This latter
outcome is only a factor 4 more likely than the emission of an inner 1s electron,
which occurs in 1.4-2.0\% of all emission processes.

In addition, table \ref{tab:ionstatepop} shows
that the emission of the inner electrons can involve the outer electrons.
The outer electrons are left in a 2s$^2$2p$^2$ configuration in 91 - 93\% of all 1s
emission processes. However, 
the final distribution over the 1s2s$^2$2p$^2$ shows deviations from the statistical
distribution of a 2:1 ratio
between the $^4$P$^e$ and the $^2$P$^e$ state, with a 5:3 ratio at 170 eV
and a ratio of 2.1:1 at a photon energy of 224 eV. In 7-9\% of the 1s
ionization processes, a change in the outer electron population occurs, mainly an
excitation of a 2s electron to 2p. This suggests that in just under 10\% of the processes,
two different electrons absorb a photon. Hence this provides a signature of
a multi-electron response to the light field, which may involve a sequential process,
whereby the first photon excites a 1s electron to 2p, followed by photoemission of one
of the 2s electrons.

The table demonstrates that all emission processes involving an outer 2s or 2p
electron decrease in magnitude with increasing photon energy. However, this pattern
changes when a 1s electron is emitted. Although the population in
the dominant residual-ion states after emission of a 1s electron
(1s2s$^2$2p$^2$ $^2$P$^e$ and $^4$P$^e$)
shows a decrease in final population with increasing photon energy, for the other
1s2s$^2$2p$^2$ states, for five out
of nine 1s2s2p$^3$ residual-ion states and all 1s2p$^4$ states, the final population
is larger at a photon energy
of 224 eV than at a photon energy of 197 eV.  This could be due to an intermediate
resonance, but it could also originate from the closer proximity of the threshold
for single-photon emission of a 1s electron at this higher photon energy.

\begin{figure}
\includegraphics[width=9cm]{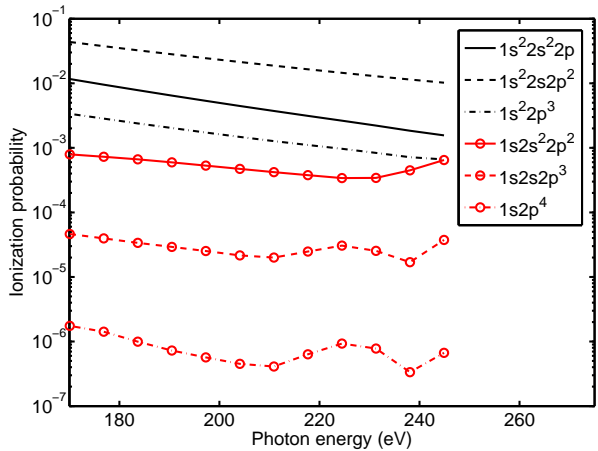}
\caption{(Colour online) Photoionization yields for residual-ion configurations of C$^+$ as a
function of photon energy at an intensity of 10$^{17}$ W/cm$^2$.}
\label{fig:carbonconf}
\end{figure}

To investigate the origin of the increase, we have carried out further calculations
of inner-shell photoemission of C atoms covering the photon-energy range between 170
and 245 eV. Outcomes of these calculations are shown in figure \ref{fig:carbonconf}. To
maintain clarity, the figure shows the probability that the C$^+$ ion is left in a
particular configuration, rather than in individual states within the configuration.
Most residual C$^+$ ions are left in the 1s$^2$2s2p$^2$ configuration, indicating that
emission of a 2s electron is the most likely photoionization process. The probability
of emission of a 2p electron is about a factor 4 times smaller at 170 eV, and this probability
decreases with increasing photon energy. The probability that the C$^+$ ion is
left in 1s$^2$2p$^3$ is over an order of magnitude smaller than the probability for
emission of a 2s electron. Emission of a 1s electron preferentially leaves the C$^+$
ion in the 1s2s$^2$2p$^2$ configuration, accounting for about 2\% of the photoionization
processes at 170 eV. This probability increases beyond 230 eV, and reaches about 5\% for
245 eV. This increase at the highest photon energies
can be ascribed to the increasing proximity of the threshold for single-photon emission
of the 1s electron.

Figure \ref{fig:carbonconf} shows that the probability of leaving the residual
C$^+$ ion in the 1s2s2p$^3$ and 1s2p$^4$ configurations are resonantly enhanced at a
photon energy of 224 eV. The populations in the 1s2s$^2$2p$^2$ configuration show
an enhancement near the threshold for single-photon emission of a 1s electron, but
this enhancement only becomes noticeable above a photon energy of 230 eV. The most
likely origin of the enhancement is the intermediate 1s2s$^2$2p$^3$ configuration
reached by photoexcitation of a 1s electron into the 2p shell. The ultra-short duration
of the pulse means that the effect of the intermediate configuration is apparent over
a broad photon energy range with very little structure. Since the effects of the
resonances are spread out over a broad photon energy range, the overall increase
in the ionization probability will be reduced, and as a consequence resonant
enhancement is only visible for the weakest photoionization channels. 

\section{Conclusions}

In conclusion, we have demonstrated the capability of RMT theory to investigate
ultra-fast inner-shell emission
processes in general multi-electron atoms. Two-photon emission of a 1s electron
from C atoms was investigated in the photon energy range between 175 and 245 eV.
At an intensity of 10$^{17}$ W/cm$^2$, two-photon emission of the 1s electron
accounts for about 2-3\% of all photoionization processes. At the two-photon level,
emission of the 1s electron is about a factor two more likely than the emission of
either a 2p or a 2s electron combined. Through examination of the final configuration
of the residual C$^+$ ion, we have furthermore observed evidence for resonant
enhancement of the 1s emission. Finally, by comparing ionization yields in different
symmetries across multiple intensities, we have determined an above-threshold
ionization yield for inner-shell photoemission of the 1s electron in C.

The determination of inner-shell processes in general multi-electron atoms poses
a number of challenges. The present calculations have been carried out using
Hartree-Fock orbitals for ground-state C. The description of the residual ion
states can be improved by expanding the initial orbital set to include
pseudo orbitals. However, the inclusion of pseudo-orbitals will lead to
significantly larger calculations. The main reason for this is that residual-ion
states are retained in the present
R-matrix calculations in order of energy. The lowest energy states will
be dominated by states with two 1s electrons. These states can involve
the Hartree-Fock 2s and 2p orbitals, but may also involve the additional
pseudo-orbitals. Hence, a significant number of states dominated by
pseudo-orbitals may have to be included before one reaches the ionic states
corresponding to emission of a
1s electron. Furthermore, the calculations require a good description of the
continuum up to high energy. This leads to much higher energies associated
with the Bloch operator. The higher energies reached in the present calculations
do not yet affect the stability of the RMT approach.

A further open question concerns the influence of non-dipole terms. The present study
assumes that the dipole approximation holds. This approximation is not unreasonable
for the interaction between the laser field and the
1s orbital. However, the approximation may not be as suitable for describing the interactions between the
$2\ell$ electrons and the laser field. The investigation of non-dipole effects would require substantial computational
development. The RMT codes themselves will need to be modified to ensure that all relevant interactions arising from
the laser field are properly taken into account. More importantly, the inner-region R-matrix codes need modification so
that higher-order transitions are calculated within the inner region, and generated in a form suitable for use
within the RMT codes. 

The authors wish to thank A.C. Brown and J.S. Parker with valuable assistance.
This work was supported by the UK EPSRC under grant no. EP/G055416/1 and the EC Initial
Training Network CORINF. This work used the
ARCHER UK National Supercomputing Service (http://www.archer.ac.uk).

\end{document}